\documentstyle[12pt]{article}
\def\ie{{\it i.e.}}

\def\av#1{\langle #1\rangle}                    
\def\pd#1#2{{\partial #1\over\partial #2}}      
\def\p2d#1#2{{\partial^2 #1\over\partial #2^2}} 
\def\td#1#2{{d #1\over d #2}}                   
\def\t2d#1#2{{d^2 #1\over d #2^2}}
\def\gtwid{\mathrel{\raise00000.3ex\hbox{$>$\kern-
.75em\lower1ex\hbox{$\sim$}}}}
\def\ltwid{\mathrel{\raise.3ex\hbox{$<$\kern-
.75em\lower1ex\hbox{$\sim$}}}}
\def\longlimit#1{\smash{\mathop{\ \longrightarrow}\limits_{#1}\ }}
\def\longsim#1{\smash{\mathop{\ \frown\!\!\smile}\limits_{#1}\ }}

\def\pe{P\!e\,}
\def\crt{c(\vec r,t)}\def\tp{\tilde p}
\def\ur{\vec{u}(\vec{r})}

\newcount\eqnum \eqnum=0  
\newcount\eqnA\eqnA=0\newcount\eqnB\eqnB=0\newcount\eqnC\eqnC=0\newcount\eqnD\eqnD=0
\def\eqnoi{\global\advance\eqnum by 1\eqno(\the\eqnum)}
\def\eqnai{\global\advance\eqnum by 1\eqno(\the\eqnum{a})}
\def\back#1{{\advance\eqnum by-#1 Eq.~(\the\eqnum)}}
\def\last{Eq.~(\the\eqnum)}                   

\def\iitem{\par\hang\textindent}
\newcount\rfignum\rfignum=0  
\def\rfigi{\medskip \global\advance\rfignum by 1\iitem{Figure \the\rfignum.~~}}

\begin{document}
\baselineskip 16pt

\thispagestyle{empty}

\begin{center}
{\large {\bf UNIVERSAL AND NON-UNIVERSAL FIRST-PASSAGE PROPERTIES OF}}
{\large {\bf PLANAR MULTIPOLE FLOWS}}

\vspace{0.3in}
\baselineskip 13pt

{\sl J. Koplik$^1$, S. Redner$^2$ and E. J. Hinch$^3$}

\vspace{0.1in}

$^1$Benjamin Levich Institute and Department of Physics\\
City College of the City University of New York, New York, NY 10031

$^2$Center for Polymer Studies and Department of Physics\\
Boston University, Boston, MA 02215

$^3$Department of Applied Mathematics and Theoretical Physics\\
Cambridge University, Silver Street, Cambridge CB3 9EW England

\end{center}

\vspace{0.3in}
\baselineskip 16pt

\centerline{\bf Abstract}

{
\narrower

\noindent  The dynamics of passive Brownian tracer particles in
steady two-dimensional potential flows between sources and sinks is
investigated.  The first-passage probability, $p(t)$, exhibits power-law
decay with a velocity-dependent exponent in radial flow and an
order-dependent exponent in multipolar flows.  For the latter, there
also occur diffusive ``echo'' shoulders and exponential decays
associated with stagnation points in the flow.  For spatially extended
dipole sinks, the spatial distribution of the collected tracer is
independent of the overall magnitude of the flow field.

\noindent P.A.C.S. Numbers: 47.55.Mh, 05.40.+j

}

\vspace{0.3in}

The motion of passive tracer in a flowing fluid is a convenient
diagnostic tool for monitoring the characteristics of the flow field, as
well as properties of the background medium.  In porous media flows,
particularly in groundwater or hydrocarbon recovery, tracer measurements
often are the only source of internal information about the system [1].
Considerable effort has focused on one-dimensional flows where tracer is
introduced at one end of the sample and the transit time distribution is
monitored upon exit at the other end.  However, in many situations, such
as fluid pumped into an ``injection'' well and extracted from one or
more ``producing'' wells, the flow is more likely to be radial, or
multipolar, rather than linear.  Our goal, in this letter, is to provide
general insights about the first-passage properties for
dynamically-neutral tracer in such flows [2].

We shall focus on steady two-dimensional flows which are generated by
multipolar configurations of sources and sinks.  These appear to
encompass many cases of possible physical relevance, as well as
illustrating a rich range of phenomenology.  Mathematically, the first
passage of passive Brownian tracer particles is determined by their
concentration, $\crt$, which satisfies the convection-diffusion equation
(CDE),
$$
\pd{\crt}{t} + \ur\cdot\vec\nabla\crt = D\nabla^2\crt,\eqnoi
$$
with the velocity field $\ur$ arising from a two-dimensional potential.
We will elucidate the features of $p(t)$, the {\em
distribution of transit times} or the {\em first-passage probability}
between source and sink, to characterize the motion of
dynamically-neutral tracer in steady flows.

First consider radial or monopolar flow, $\ur =u_0\hat r/r$.
Non-universal features arise due to the competition between the flow and
the centrifugal term in the Laplacian, which both vary as $1/r$.
Consequently, the form of the CDE in two dimensions ($d=2$) with $u_0$
finite is the same as that of a system with no drift but with $d\ne 2$.
Thus changing the amplitude $u_0$ is equivalent to changing the spatial
dimension, leading to non-universal first-passage
properties as a function of the drift.

To illustrate these features, we have solved for  $p(t)$
in a circular geometry with an inner absorbing radius $a$, infinite
outer radius, and an initial ring of tracer particles which are released
at $r=r_0$.  This first passage probability coincides with the radial flux at
$r=a$, \ie, $p(t)=-D\partial c(r,t)/\partial r+u(r)c(r,t)$.
Using standard Green's
function methods, the Laplace transform,
$\tilde p(s)\equiv \int_0^\infty p(t)\, e^{-st}\,dt\/$, is [2]
$$
\tp (s)=\left({a\over{x_0}}\right)^\nu {K_\nu(x_0\sqrt{s/D})
\over {K_\nu(a\sqrt{s/D})}}.
\eqnoi
$$

where $K_\nu$ is the modified Bessel function of the second kind, and
$\nu=u_0/2D\equiv\pe /2$, where $\pe$ is the P\'eclet number.  The
long-time behavior is determined from the small-$s$ expansion of $\tp$,
which generically has the form, $\tp (s)\sim \tp (0) - a s^{\alpha} + b
s^1 + \ldots$.  By construction, the leading term $\tp(s=0)=
\int_0^\infty p(t)\,dt\equiv E(r_0)$ is just the probability that
a tracer particle which starts at $r=r_0$ {\it eventually} reaches the
absorbing circle.  Time-dependent properties can be inferred from the
correction terms in \last.  When $\alpha<1$, then the mean first-passage
time to the absorbing circle, $\av{t}$, diverges and the first-passage
probability in real time has a power law tail which varies as
$t^{-(\alpha+1)}$.  However, when $\alpha$ reaches unity, the
coefficient of $-s$ equals $\av{t}$.

For outward flow, the small-$s$ expansion of $p(s)$ is
$$
\tp (s)\sim \left({a\over x_0}\right)^{2\nu}
\left(1 - {\Gamma(1-\nu)\over{\Gamma(1+\nu)}}
(r_0^{2\nu}-a^{2\nu}) \left({s\over 4D}\right)^\nu +\ldots\right),
\eqnoi
$$
Thus $E(r_0)=(a/r_0)^{2\nu}<1$ and some tracer escapes to infinity.
This reflects the equivalence to a purely diffusive system of spatial
dimension $d>2$, for which the eventual return probability is less than
unity [3].  From the correction term, we deduce that $\av{t}\to\infty$
for $0\leq\nu<1$, while for $\nu\ge 1$, $\av{t}$ is finite.  Since
$E(r_0)<1$, $\av{t}$ in the case $\nu>0$ is conditional
on those particles which actually hit the absorber.  Note that for
sufficiently large outflow, a tracer particle must reach the
absorber in a finite time if it is to be absorbed at all.

For inflow ($\nu<0$), the first passage probability has the small-$s$
expansion
$$
\tp (s)\sim 1 - {\Gamma(1-\mu)\over{\Gamma(1+\mu)}}
(r_0^{2\mu}-a^{2\mu}) \left({s\over 4D}\right)^\mu  +\ldots,
\eqnoi
$$
where $\mu\equiv-\nu>0$.  Thus tracer hits the absorbing circle with
certainty, while $p(t)$ is non-universal and varies as $t^{-(1+\mu)}$.
Additionally for $\mu<1$, $\av{t}$ is infinite, while for $\mu\geq 1$,
$\av{t}$ is finite.  Even though the tracer is guaranteed to reach the
absorber for any inflow (including zero), the first-passage time
is finite only if the drift is sufficiently strong.

We now discuss tracer motion in a fully two-dimensional dipole flow
field due to a point source and sink at
$(0,\mp a)$ within a circle of radius $R$, with total flux
$\pm Q$.  The  velocity field may be written as
$$
\vec{u}=\nabla\phi =(\pd{\psi}{y},-\pd{\psi}{x}).\eqnoi
$$
Here the velocity potential $\phi$ and streamfunction $\psi$ are the real
and imaginary parts of the complex potential
$$
\Psi =\phi +i\psi = {Q\over 2\pi} \ln{(z+a)(z+R^2/a)\over (z-a)(z-R^2/a)},
\eqnoi
$$
where $\pm R^2/a$ are the image locations to impose vanishing
normal velocity on the boundary, and $z=x+iy$.
We assume that fluid and tracer are
emitted uniformly in angle about the source.

In the limit of pure convection in an infinite plane ($R\to\infty$), the
tracer motion is determined by the velocity field,
$d\vec{r}/dt=\vec{u}(\vec{r}(t))$.  Since the second equality in \back1
represents a Hamiltonian system, we have applied a canonical
transformation technique to compute particle trajectories analytically
[4], and thereby determine that the transit time from source to sink for
a particle emitted at polar angle $\theta$ with respect to the source is
$ t(\theta )=2\csc^2{\theta}\ [1-\theta\cot{\theta}\ ]$; see also [5].
Although the motion is deterministic, a distribution in transit times
results from the distribution in initial angle.  Since all streamlines
lead from source to sink, there is a one-to-one equivalence between the
angular and time dependence of the first-passage probability.  This
leads to
$$
p(t)=p(\theta )\ \left| \td{\theta}{t}\right|
\longlimit{\theta\to\pi} {\scriptstyle {1\over 3}} (2\pi)^{-2/3}\cdot t^{-4/3},
\eqnoi
$$
due to tracer particles which initially move away from the sink on
distant dipole streamlines.  It is
this aspect of the distribution which explores the global structure of
the flow field.  Note that the mean transit time,
$\av{t}=\int_0^{\infty}dt\, t\, p(t)$, diverges.

To incorporate molecular diffusion, we resort to numerical simulation,
and have employed several complementary methods.  The simplest is
grid-free Monte Carlo time-stepping using individual random walkers,
which works best at high P\'{e}clet numbers.  In time $\Delta t$, a
walker is displaced by $\Delta\vec{r} =\vec{u}(\vec{r})\Delta
t+\hat{n}(4\Delta t/Pe)^{1/2}$, where $\hat{n}$ is a unit vector of
random orientation.  Figure~1(a) gives a typical result for $\pe=2700$ with
$N=500,000$ random walkers.  There is an early-time peak, followed by
the anticipated $t^{-4/3}$ power-law decay, and then a noisy exponential
region dominated by diffusion.  To clarify the latter domain, we have
devised a lattice ``probability propagation'' algorithm.  A probability
element at $(i,j)$ translates through a distance $\Delta\vec{r}=
\vec{u}(\vec{r})\Delta t$ in a time $\Delta t$ to an off-lattice
position $(x,y)$, and is then redistributed among the 5-site nearest
lattice neighborhood of the target position.  The redistribution rule is
chosen to ensure that the average displacement of the 5-site group
remains equal to $\Delta\vec{r}$ and that diffusion is spatially
constant by imposing a constant fluctuation in the displacement of this
group.  While this method has no statistical fluctuation, the simple
form of the redistribution rule restricts the method to low flow rates.

The probability propagation results are equivalent to standard
finite-difference methods, but programming is extremely simple.  A
typical result for $p(t)$ (Fig.~1(b)), exhibits the following four
generic features: an early-arrival regime, a power-law decay, a
``diffusive echo'' shoulder (whose resolution in simulations is
dependent on the absence of statistical fluctuations), and an ultimate
exponential decay.  The shoulder stems from particles which reflect from
the boundary before reaching the sink.  Because of the low P\'{e}clet
number involved, non-trivial asymptotic exponent estimation and
extrapolation [2] are required to verify the exponent in the power-law
region.  Since $p(t)\sim 1/(t\ln^2 t)$ in the diffusive limit [3] and
$p(t)\sim t^{-4/3}$ as $\pe\to\infty$, it seems unlikely that a
simulation method which cannot be applied at large $\pe$ will be able to
resolve these two limiting behaviors cleanly.

The decay at the longest times may be correlated with {\em
stagnation points} in the flow field which generically lead to an
exponential decay in the transit-time distribution [6].  Without loss of
generality, consider a stagnation point at the origin with the local
velocity $\vec{u}=d\vec{r}/dt= (-Gx,Gy)$.  If the stagnation point is
approached along the $x$-axis, then trajectories which pass near the
origin and then escape have the form $\vec{r}(t)=(x_0 e^{-Gt},y_0
e^{Gt})$ with $y_0 \ll x_0$.  The time $T$ spent near the stagnation
point can be defined by requiring that $\dot{y}\sim U_0$, where $U_0$ is
some characteristic O(1) velocity.  Since $\dot{y}\sim y_0 G e^{GT}$,
this gives $T\sim G^{-1} \ln{(U_0/y_0 G)}$.  Typically, for trajectories
which pass near a stagnation point, the transit time between source and
sink is dominated by this value of $T$.  The corresponding distribution
of transit times may now be obtained by accounting for the distribution
of initial positions $y_0$.  Consequently,
$$
p(T)=p_0(y_0)\left|{dy_0\over dT}\right|
= p_0({U\over G}e^{-GT}) \left| -Ue^{-GT}\right|
\longsim{T\to\infty} p_0(0) e^{-GT}. \eqnoi
$$
Thus the transit time distribution depends on the local shear rate near
the stagnation point, but not on details of the initial spatial
distribution of tracer $p_0$.  In the present example, applying
Eqs.~(5), (6), and (8) to the stagnation points at $(\pm R,0)$ gives an
exponential decay rate with $G\propto R^{-3}$.  The decay rate observed
in simulations will typically be the smaller of this value and the
diffusive decay rate, which is proportional to $R^{-2}$ for the dipole case.

For a source-sink of arbitrary multipolarity, the exponent of the
power-law tail in $p(t)$ depends on the multipole moment.  This exponent
can be obtained by the following simple argument.  For a $2^N$-pole, the
leading behavior of the complex potential is $\Psi\sim z^{-N}$, so that
the corresponding streamlines are $\psi={\rm Im}\,\Psi \sim
r^{-N}\sin{N\theta}$, for suitable orientation of the axes.  In the
convective limit, particle trajectories are defined by $\psi={\rm
const.}$.  This constraint implies that $\psi (r,\theta )=\psi(R,\pi
/2N)=R^{-N}$, or $r^N=R^N\sin{N\theta}$, where $R$ is the maximum
distance from the origin on the streamline.  The angular velocity on
such a streamline is

$$
\td{\theta}{t}=-{1\over r}\pd{\psi}{r} \sim Nr^{N-2}\sin{N\theta}
 \sim R^{-N-2}\sin^{-2/N}{(N\theta )}.\eqnoi
$$
The transit time on this trajectory is the time required for
$\theta$ to vary between 0 and $\pi /N$.  Using the above
approximation for the angular velocity, the transit time scales as
$ t=\int_0^{\pi /N}d\theta \,(d\theta /dt)^{-1} \sim R^{N+2}$.
To determine the transit time distribution, we relate the value
of $R$ to the initial angle of emission $\theta_0$ at the source.
Suppose the source is at the origin and oriented so that the
long-excursion streamlines are associated with $\theta_0\to 0$.  For
$r\to\epsilon$, $\psi =R^{-N} \to\epsilon^{-N}\sin{\theta_0}\to
\epsilon^{-N}\theta_0$, or $\theta_0\sim R^{-N}$.  Thus the transit time
probability distribution is
$$
p(t)=p(\theta_0)\left|\td{\theta_0}{t}\right|={1\over
2\pi}\left|\td{\theta_0}{R}
\td{R}{t}\right|\sim R^{-2N-2}\sim t^{-{2N+2\over N+2}}.
\eqnoi
$$
While these predictions have been verified numerically in the convective
limit,  there are as-yet unexplained diffusive features in
probability propagation simulations.  For example, the form of $p(t)$
for a quadrupole consisting of charges $(-Q,2Q,-Q)$ is quite different
from that of an ``inverted'' quadrupole $(Q,-2Q,Q)$.

Another important diagnostic is the distribution of {\em where} on the
sink the tracer is collected.  In the {\em dipole} case, with passive
tracer released in steady potential flow between a source and a single
spatially extended equipotential sink, we demonstrate that the
time-integrated tracer flux distribution at a given point on the sink is
proportional to the incoming fluid velocity at this point.  By
linearity, this implies that the spatial distribution of the collected
tracer is {\em independent} of the overall magnitude of the flow.  This
establishes a useful general equivalence between tracer distributions
for pure diffusion and in combined convection and diffusion.

To prove the theorem, we use the complex potential as the conformal
mapping to transform the flow domain from $(x,y)$ into $(\phi ,\psi )$.
In these variables the CDE becomes
$$
|\vec{u}(\phi ,\psi )|^{-2} \pd{c}{t}+\pd{c}{\phi}={1\over\pe}\left(
\p2d{c}{\phi}+ \p2d{c}{\psi} \right) .
\eqnoi
$$
This transformation to linear flow introduces a spatially dependent
time step which is singular as $(0,i\pi)$ is approached.  In the $(\phi
,\psi)$ coordinate system, the source has co-ordinate $\phi =\phi_0$.
For the initial condition of a delta-function pulse of tracer injected
at the source, then in the equation of motion for $\xi(\phi ,\psi)=
\int_0^\infty dt\, c(t,\phi,\psi)$, the time derivative term
integrates to zero, yielding
$$
\pe\pd{\xi}{\phi}=\p2d{\xi}{\phi}+\p2d{\xi}{\psi},\eqnoi
$$
with boundary conditions $(1-\pe^{-1}\partial /\partial\phi )\xi =K$,
a $\pe$-independent constant, at $\phi =\phi_0$, and $\xi =0$ at
$\phi=\phi_1$.  The solution of \last\ is
$\xi=K(1-e^{\pe (\phi -\phi_1)})$, and
the time-integrated tracer flux arriving at angular position $\psi$ on
the sink, $p(\psi)$, is
$$
p(\psi ) = -\pe^{-1}\int_0^\infty dt\, \pd{c}{n}(\phi_1,\psi,t)
= -\pe^{-1}\pd{\xi}{n}(\phi_1,\psi )
= -\pe^{-1}\left. \pd{\xi}{\phi}\pd{\phi}{n}\right|_{(\phi_1,\psi )} \eqnoi
$$
where $\hat{n}$ is the unit normal to the sink.  In the last
expression, the first factor is a constant, and the second is simply the
normal velocity of fluid at the sink.  Hence $p(\psi )\propto u_n(\psi)$
as claimed.

Notice that the theorem also holds in the pure diffusion limit, as can
be seen by taking $\pe\to 0$ either in the equation for $\xi$ or in its
solution.  In this case, the appropriate statement is that $p(\psi)$ is
constant for a constant arclength of the sink.  Conversely in the limit
of no diffusion the theorem is obvious, because then tracer particles
remain on their initial streamline, and the tracer flux is simply
proportional to the fluid flux.  The nontrivial implication is that the
local integrated flux to a simply-connected sink is independent of the
P\'eclet number, although the transient flux is P\'eclet number
dependent.

In summary, we have investigated the first-passage probability $p(t)$ of
dynamically-neutral tracer in two-dimensional potential flows.  For
radial flow, convection and diffusion are of the same order leading to
non-universal first-passage characteristics.  In general multipole
flows, the exponent of $p(t)$ depends on the multipole order.  The
effects of diffusion appear to be subdominant for the dipole but
relevant for the quadrupole, as qualitative features of $p(t)$ depend on
the sense of the quadrupolar flow when diffusion is present.  The
influence of stagnation points on the asymptotic properties of $p(t)$
was determined.  Finally, the time-integrated flux to a fixed arclength
of sink is independent of the overall magnitude of the flow field.

\newpage
\centerline{\bf Acknowledgements}

We thank Jean Pierre Hulin and Michael Stephen for useful discussions.
This research was partially supported by DOE grant DE-FG02-93-ER14327 to
JK and NSF grant DMR-9219845 to SR.

\vspace{0.3in}

\centerline{\bf References}

\begin{enumerate}

\item J. Bear, {\sl Dynamics of Fluids in Porous Media}, (Elsevier, Amsterdam,
1971);
F. A. L. Dullien, {\sl Porous Media: Structure and Fluid Transport},
2nd ed. (Academic, London, 1991);
E. Guyon, J.-P. Nadal and Y. Pomeau, eds., {\sl Disorder and Mixing},
(Kluwer, Dordrecht, 1988).

\item For full details, see J. Koplik, S. Redner, and E. J. Hinch,
``Tracer Dispersion in Planar Multipole Flows'', to be published.

\item See, {\it e.\ g.}, G. H. Weiss and R. J. Rubin, {\sl Adv.\ Chem.\ Phys.}
{\bf 52}, 363, (1983) and references therein.

\item H. Goldstein, {\sl Classical Mechanics}, 2nd ed., (Addison-Wesley,
Reading MA, 1980).

\item L. Mittag and M. J. Stephen, ``Motion and Diffusion of a Passive Scalar
in a Two Dimensional Fluid'', to be published.

\item P. Kurowski, I. Ippolito, J. P. Hulin, J. Koplik and E. J. Hinch,
{\sl Phys.\ Fluids} {\bf 6}, 108 (1994).

\end{enumerate}
\vspace{0.3in}
\parindent=0.8in
\centerline{\bf Figure Caption}

\rfigi
First-passage probability distribution in combined dipolar flow and
molecular diffusion within a circle of radius $R=400$ with source and
sink at $x=\mp 20$, obtained by (a) following the motion of individual
random walks for $\pe = 2700$ and (b) probability propagation for
P\'eclet numbers $\pe=1.6$, 0.8, 0.4, and 0.2 (tallest to shortest peak,
respectively).  The inset in (a) shows the streamlines.  Also in (a),
the abscissa is the dimensionless time $t\to t\cdot(2\pi a^2/Q)$.

\end{document}